%% file: JB_Curac_2026.tex
\documentclass[USenglish,twocolumn]{article}
\usepackage[utf8]{inputenc}
\usepackage[big,online]{dgruyter}
\usepackage[sort&compress,square,numbers]{natbib}
\usepackage{times}
\usepackage{graphicx}
\usepackage{subcaption}
\usepackage{mathtools}
\usepackage{xcolor}

\begin{document}

  \DOI{10.1515/}
  \openaccess
  \pagenumbering{gobble}

\title{Investigating the trackability of landmarks in NBI and WLI}
\runningtitle{Trackability of landmarks in NBI and WLI}

\author*[1]{Jonas Below}
\runningauthor{J. Below et al.}
\author[2]{Piermarco Gerli}

\author[3]{Anna Jötten}

\author[3]{Dennis Eggert}

\author[4]{Axelle Felicio-Briegel}

\author[4]{Veronika Volgger}

\author[3]{Christian Betz}

\author[2]{Alexander Schlaefer}

\affil[1]{\protect\raggedright 
  Institute of Medical
Technology and Intelligent Systems, Hamburg University of
Technology, Am Schwarzenberg-Campus 3, Hamburg, Germany, e-mail: jonas.below@tuhh.de}

\affil[2]{\protect\raggedright
   Institute of Medical
Technology and Intelligent Systems, Hamburg University of
Technology, Hamburg, Germany}

\affil[3]{\protect\raggedright 
   Department of Otorhinolaryngology, Head and
Neck Surgery and Oncology, University Medical Center
Hamburg-Eppendorf, Martinistraße 52, Hamburg, Germany}

\affil[4]{\protect\raggedright 
LMU Klinikum, Marchioninistr. 15,
81377 München, Germany}

\abstract{Reliable visual tracking of anatomical landmarks in endoscopic video is a key prerequisite for computer-assisted interventions in laryngeal surgery. While Narrow Band Imaging (NBI) is clinically established for enhancing vascular contrast, its effect on automated tracking relative to conventional White Light Imaging (WLI) remains unexplored. We present a quantitative comparison of landmark trackability in NBI and WLI using a forward–backward error framework applied to 61 endoscopic videos from 39 patients. Two regions of interest are evaluated: the vocal folds (VF) and mucosa with a prominent vascular pattern (MPVP). We compare a classical feature-based tracker (CSRT) against learning-based point trackers (CoTracker3-Online and TrackOnR) across both modalities. NBI consistently reduced the forward–backward error in both regions, reaching statistical significance for the vocal folds, while failure rates stayed low. These findings indicate that NBI can improve tracking robustness in laryngeal endoscopy, with implications for motion analysis, navigation, and robotic assistance.}


\maketitle

\section{Introduction} 
\input{01_introduction}
\section{Methods}
\input{02_methods}
\section{Results}
\input{03_results}
\section{Discussion}
\input{04_discussion}

\textsf{\textbf{Author Statement}}\\
This research was co-funded by the uNBIased project (DKH, grant 70116520), by the HyperTUM project (BMBF, grant 16SV9567), and by the MARLOC project (DFG,  grant SCHL 1844-10-1). Conflict of interest: Authors state no conflict of interest. Informed consent: Informed consent has been obtained from all individuals included in this study. Ethical approval: The research related to human use complies with all the relevant national regulations, institutional policies and was performed in accordance with the tenets of the Helsinki Declaration, and has been approved by the authors' institutional review board or equivalent committee.

\bibliographystyle{plainnat}
\bibliography{references}
\end{document}

%% file: 01_introduction.tex
Reliable markerless tracking of natural landmarks in endoscopic videos is an important prerequisite for many computer-assisted interventions and analysis tasks, including motion compensation~\cite{motion_compensation} and navigation~\cite{navigation}. In laryngeal endoscopy, robust tracking of anatomical landmarks is particularly challenging due to frequent camera motion, tissue deformation, illumination changes, and partial occlusions. Accurate temporal tracking of structures such as the vocal folds could nonetheless support downstream applications such as automated screening workflows and robotic assistance systems.

Besides conventional White Light Imaging (WLI), Narrow Band Imaging (NBI) is widely used in clinical practice to enhance the visualization of vascular structures and mucosal patterns. Soft tissue tracking in surgical scenes has been studied extensively for WLI~\cite{surgt}, but the performance of modern trackers on NBI remains unexplored. From a computer vision perspective, the enhanced vascular texture in NBI could improve the temporal stability of visual features and thereby tracking robustness, while the altered image appearance and stronger local contrast may instead degrade certain methods. It thus remains unclear whether NBI generally improves trackability over WLI, and whether any benefit depends on the tracked anatomical region or the tracking algorithm.

In this work, we present a quantitative comparison of landmark trackability in NBI and WLI laryngeal endoscopic videos. We evaluate tracking consistency on two regions of interest (ROIs) using the forward--backward tracking error across 61 videos from 39 patients, employing two state-of-the-art point trackers and one baseline box tracker.



%% file: 02_methods.tex
\subsection{Data processing}
The evaluation data were acquired as described in \cite{bhattacharya2022learning} and extended from 14 to 39 patients, yielding 61 endoscopic videos. The study was approved by the local ethics committee (Project number 44-15, Ludwig Maximilian University of Munich). Each video contained both WLI and NBI recordings, which were manually identified and cropped into modality-specific segments. Segments were screened, and only those longer than 2\,s and suitable for tracking were retained; those with extensive blur (e.g., from saliva) or rapid motion (e.g., coughing) were excluded. Each retained segment was then cut into overlapping subvideos of 2, 4, and 6\,s, with a new set generated every 4\,s along the timeline.

\subsection{Experimental setup}
The first frame of each subvideo was annotated for two regions of interest (ROIs): the vocal-fold region (VF), including immediately adjacent tissue, and the mucosa characterized by a prominent vascular pattern (MPVP), a manually selected mucosal area with prominently visible vessels. As MPVP regions were not a primary focus during acquisition, they were more often affected by motion, defocus, or partial loss of visibility.

To initialize the tracker, each annotated ROI was subdivided into uniformly distributed, axis-aligned square bounding boxes covering the region, each treated as an independent tracking candidate. Point-based trackers tracked only the centroid of each candidate, whereas the box-based tracker tracked its full spatial extent.

To obtain comparable candidate densities across videos, we defined target candidate counts ($\tilde{n}$) per ROI with a predefined tolerance interval, since packing an exact number of square boxes into an arbitrary polygon is often computationally difficult or geometrically infeasible. A solution was accepted when the resulting box count ($n$) fell within $[n_{\min}, n_{\max}]$. For $\tilde{n}=10$, $100$, and $300$, the acceptable intervals were $[6,14]$, $[70,130]$, and $[200,400]$, respectively.

\begin{figure}
    \centering
    \includegraphics[width=0.8\linewidth]{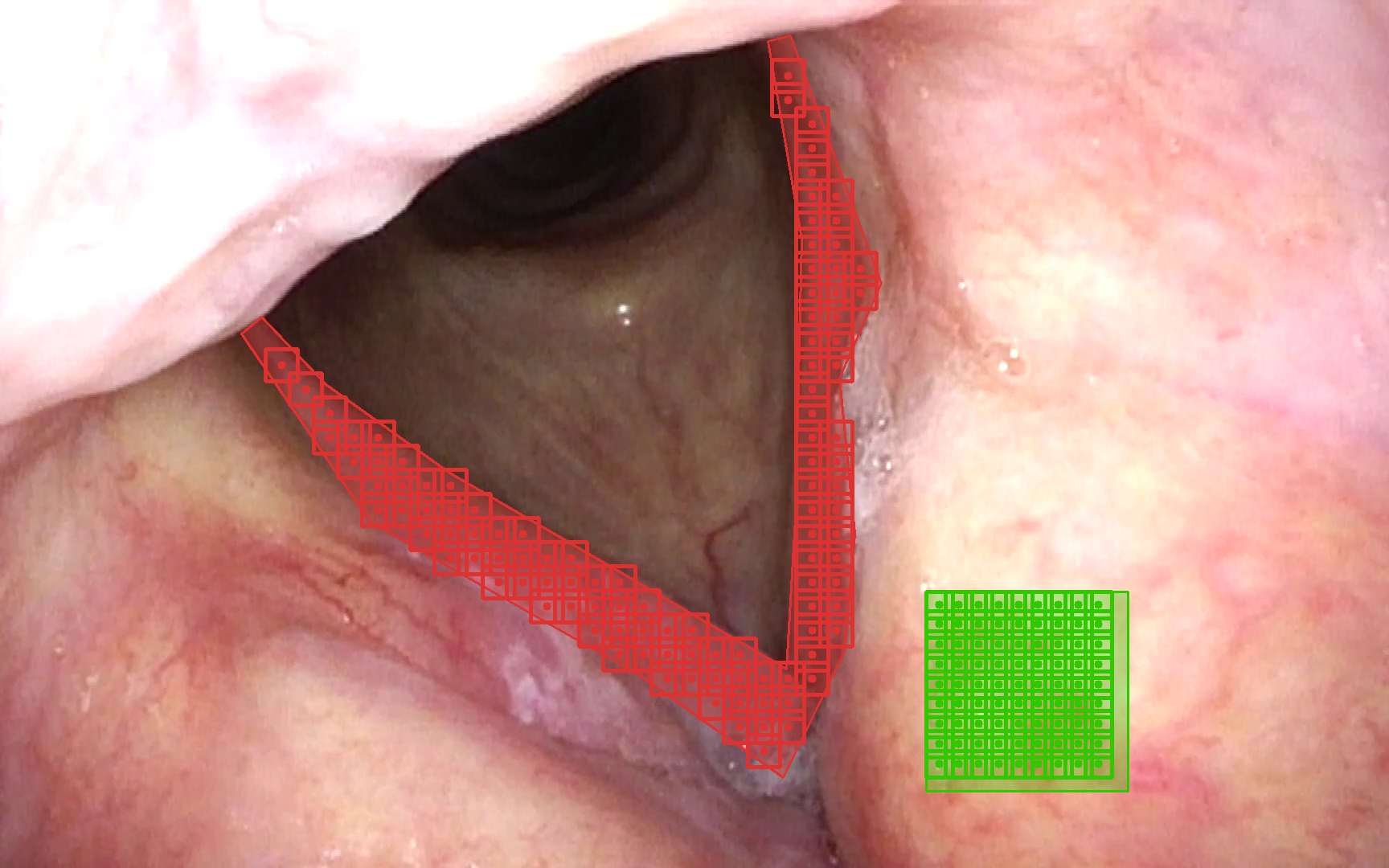}
    \caption{Example placement of $\tilde{n}=100$ tracking candidates within manually annotated VF (red) and MPVP ROIs (green). For point-based trackers, the centroid of each box was used as the tracking point.}
    \label{fig:raster_example}
\end{figure}

An ROI was evaluated only if it remained fully visible in the subvideo's final frame; ROIs occluded in the final frame were excluded, while temporary occlusions in earlier frames were permitted.

\subsection{Forward--Backward Tracking Consistency and Trackability Estimation}

To assess tracking reliability, each tracker instance was run in both forward and backward temporal directions, producing two trajectories of identical shape $(T, n)$, where $T$ is the number of frames. The forward trajectory ($\mathrm{fwd}$) was initialized from all candidates at frame $t = 0$ and propagated sequentially to frame $t = T-1$. For point-based trackers, each candidate was represented by its centroid. For box-based trackers, the full bounding box was used. The backward trajectory ($\mathrm{bwd}$) was then initialized from the terminal forward position at $t = T-1$ and propagated back to $t=0$. For evaluation, all analyses used centroid trajectories regardless of tracker type.

Tracking reliability was quantified via a forward--backward error (FBE). For each candidate and frame $t \in \{0, \dots, T-1\}$, the per-frame error $e_t$ was the Euclidean distance between the forward and backward centroids:

\begin{equation}
e_{t} =
\left\lVert
\mathbf{c}^{\mathrm{fwd}}_{t}
-
\mathbf{c}^{\mathrm{bwd}}_{t}
\right\rVert_2 .
\end{equation}

This measures the tracker's temporal self-consistency: a reliable tracker should return to its original position after tracking forward and then backward, whereas large forward--backward discrepancies indicate uncertainty, drift, or temporal instability.

For each candidate, the mean forward--backward error $\mathrm{mFBE_{tc}}$ was computed by averaging $e_t$ over all frames except the last, which is excluded since $e_{T-1}=0$ by definition. Lower values indicate higher temporal consistency. The $\mathrm{mFBE_{tc}}$ values were averaged over all candidates within a subvideo and ROI, then aggregated at the video level to obtain $\mathrm{mFBE_v}$. Modality-level results were computed from these video-level values, preventing videos with more subvideos or candidates from dominating the overall $\mathrm{mFBE}$.

From these local consistency measurements, we define \emph{trackability} as the spatial distribution of temporal consistency across the tracked ROI: regions with consistently low FBE are highly trackable, while those with large forward--backward discrepancies are poorly trackable. The trackability map was obtained by spatially sampling the ROI with the predefined bounding boxes and assigning each its $\mathrm{mFBE_{tc}}$ value. Its spatial resolution is set by the spacing and size of the tracking boxes: finer tiling increases resolution at higher computational cost, while coarser tiling yields smoother but less detailed estimates.

As a second metric, we computed the Failure Rate (FR) from the Intersection-over-Union (IoU) between the initial candidate box and the box recovered after the forward--backward cycle. For point-based trackers, the tracked centroid was converted back into a box using the original candidate dimensions, applying the same IoU criterion to both tracker types. A candidate was labeled as failed when the IoU between its initial box at $t=0$ and the recovered backward box at $t=0$ was zero. The video-level failure rate was the mean of the binary failure indicators over all $n$ candidates:

\begin{equation}
\mathrm{FR_v}
=
\frac{1}{n}
\sum_{i=1}^{n}
\mathbf{1}_{\mathrm{IoU}_i = 0}
\end{equation}

The overall FR was aggregated analogously to the overall mFBE.

Collectively, this forward--backward framework provides a self-supervised estimate of local tracking consistency without dense frame-wise annotations. Because it relies solely on trajectory consistency under temporal reversal, it applies uniformly to point and box trackers across large video datasets with minimal manual intervention.

%% file: 03_results.tex
\begin{table}[t]
\centering
\caption{mFBE in pixels ($\downarrow$) for MPVP and VF, including the p-value of the modality comparison.}
\label{tab:mfbe-scores}
\begin{tabular}{lccc}
\toprule
Tracker & NBI & WLI & p-value $\downarrow$\\
\midrule

\multicolumn{4}{c}{\textbf{MPVP}} \\
\midrule
CoTracker3 & $4.91 \pm 11.43$ & $16.68 \pm 47.33$ & $0.123$ \\
TrackOnR           & $2.62 \pm 2.84$  & $23.86 \pm 101.72$ & $0.159$ \\
CSRT              & $298.60 \pm 164.27$ & $312.32 \pm 198.95$ & $0.395$ \\

\midrule
\multicolumn{4}{c}{\textbf{Vocal Folds}} \\
\midrule
CoTracker3 & $4.22 \pm 2.59$ & $7.33 \pm 7.87$ & $0.013$ \\
TrackOnR           & $3.81 \pm 2.89$ & $8.06 \pm 18.36$ & $0.084$ \\
CSRT              & $157.13 \pm 116.26$ & $226.45 \pm 145.04$ & $0.013$ \\

\bottomrule
\end{tabular}
\end{table}

\begin{table}[t]
\centering
\caption{FR in \% ($\downarrow$) for MPVP and VF, including the p-value of the modality comparison.}
\label{tab:fr-scores}
\begin{tabular}{lccc}
\toprule
Tracker & NBI & WLI & p-value $\downarrow$ \\
\midrule

\multicolumn{4}{c}{\textbf{MPVP}} \\
\midrule
CoTracker3 & $2.4~\% \pm 9.6~\%$ & $2.1~\% \pm 4.6~\%$ & $0.560$ \\
TrackOnR           & $0.6\% \pm 1.4\%$                  & $4.6\% \pm 20.3\%$                  &  $0.172$     \\
CSRT              & $82.8~\% \pm 12.9~\%$ & $82.9~\% \pm 15.5~\%$ & $0.497$ \\

\midrule
\multicolumn{4}{c}{\textbf{Vocal Folds}} \\
\midrule
CoTracker3 & $1.4~\% \pm 4.7~\%$ & $2.7~\% \pm 5.1~\%$ & $0.126$ \\
TrackOnR           & $0.9\% \pm 1.5\%$                  & $2.6\% \pm 6.0\%$                  & $0.053$       \\
CSRT              & $61.0~\% \pm 24.0~\%$ & $78.1~\% \pm 15.7~\%$ & $0.001$ \\ 
\bottomrule
\end{tabular}
\end{table}

\begin{figure}[htbp]
    \centering

    \begin{subfigure}[t]{0.48\columnwidth}
        \centering
        \includegraphics[width=\linewidth]{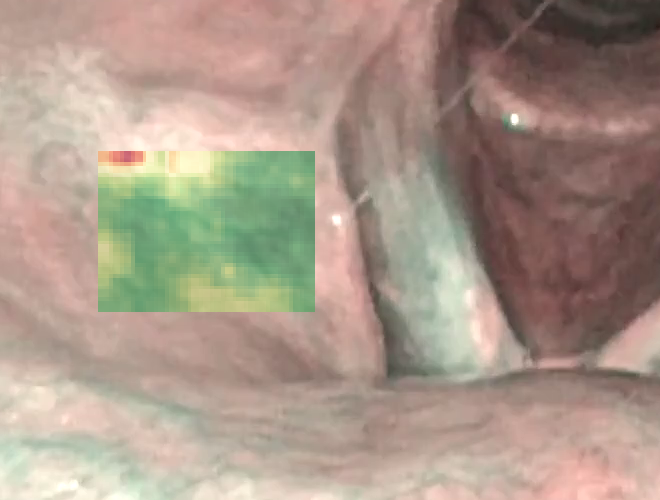}
        \caption{MPVP ROI in NBI, vestibular fold; $\mathrm{mFBE_{tc}} \in [1,5]$ px.}
        \label{fig:hvdr_nbi}
    \end{subfigure}
    \hfill
    \begin{subfigure}[t]{0.48\columnwidth}
        \centering
        \includegraphics[width=\linewidth]{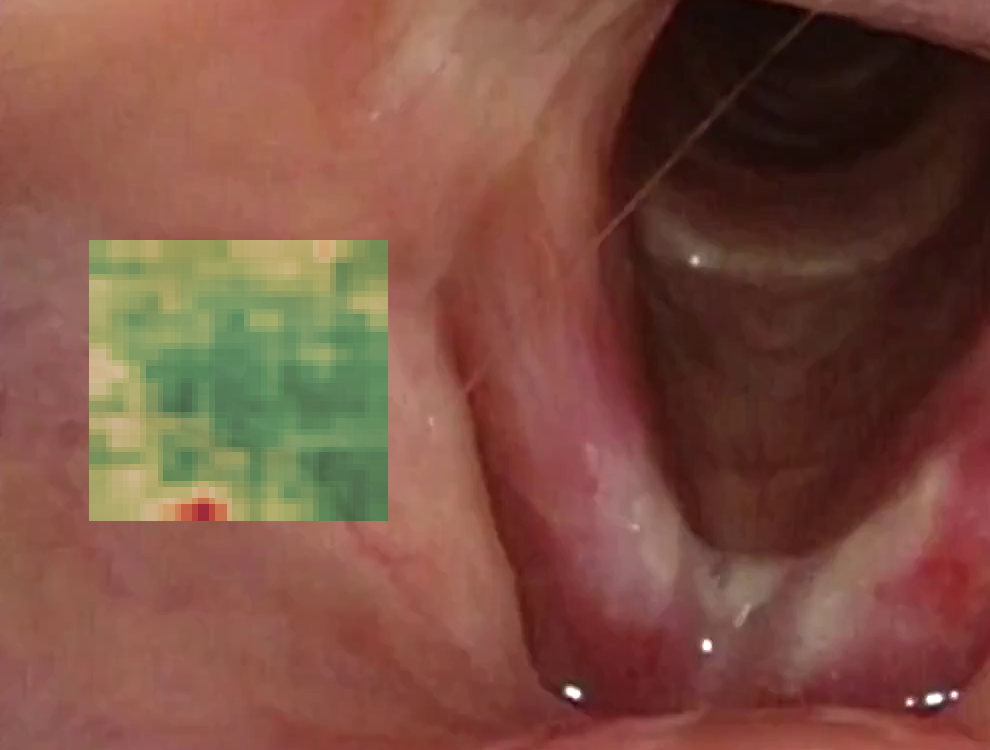}
        \caption{MPVP ROI in WLI, vestibular fold; $\mathrm{mFBE_{tc}} \in [1,7]$ px.}
        \label{fig:hvdr_wli}
    \end{subfigure}

    \vspace{0.5em}

    \begin{subfigure}[t]{0.48\columnwidth}
        \centering
        \includegraphics[width=\linewidth]{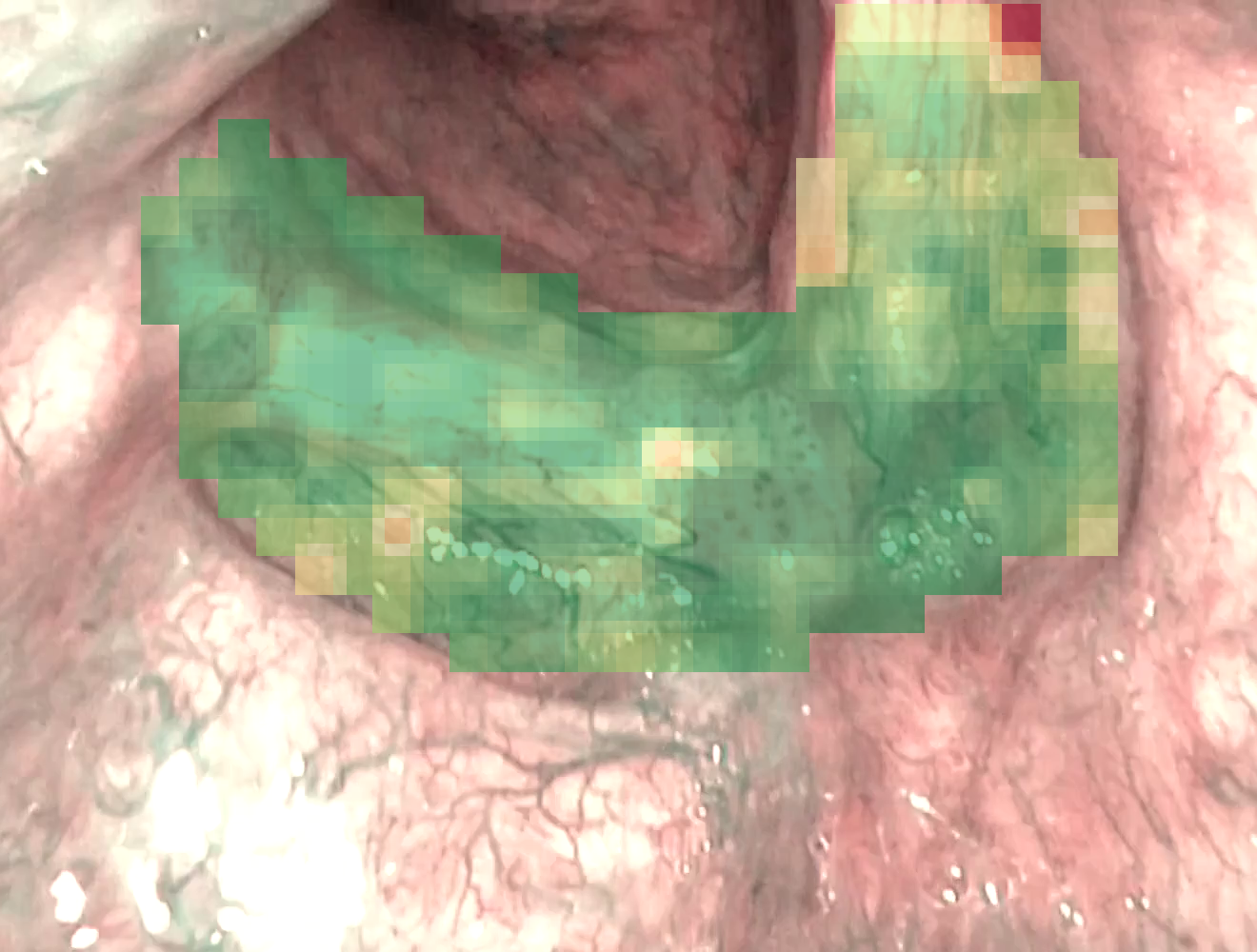}
        \caption{VF ROI in NBI; $\mathrm{mFBE_{tc}} \in [1,7]$ px.}
        \label{fig:vf_nbi}
    \end{subfigure}
    \hfill
    \begin{subfigure}[t]{0.48\columnwidth}
        \centering
        \includegraphics[width=\linewidth]{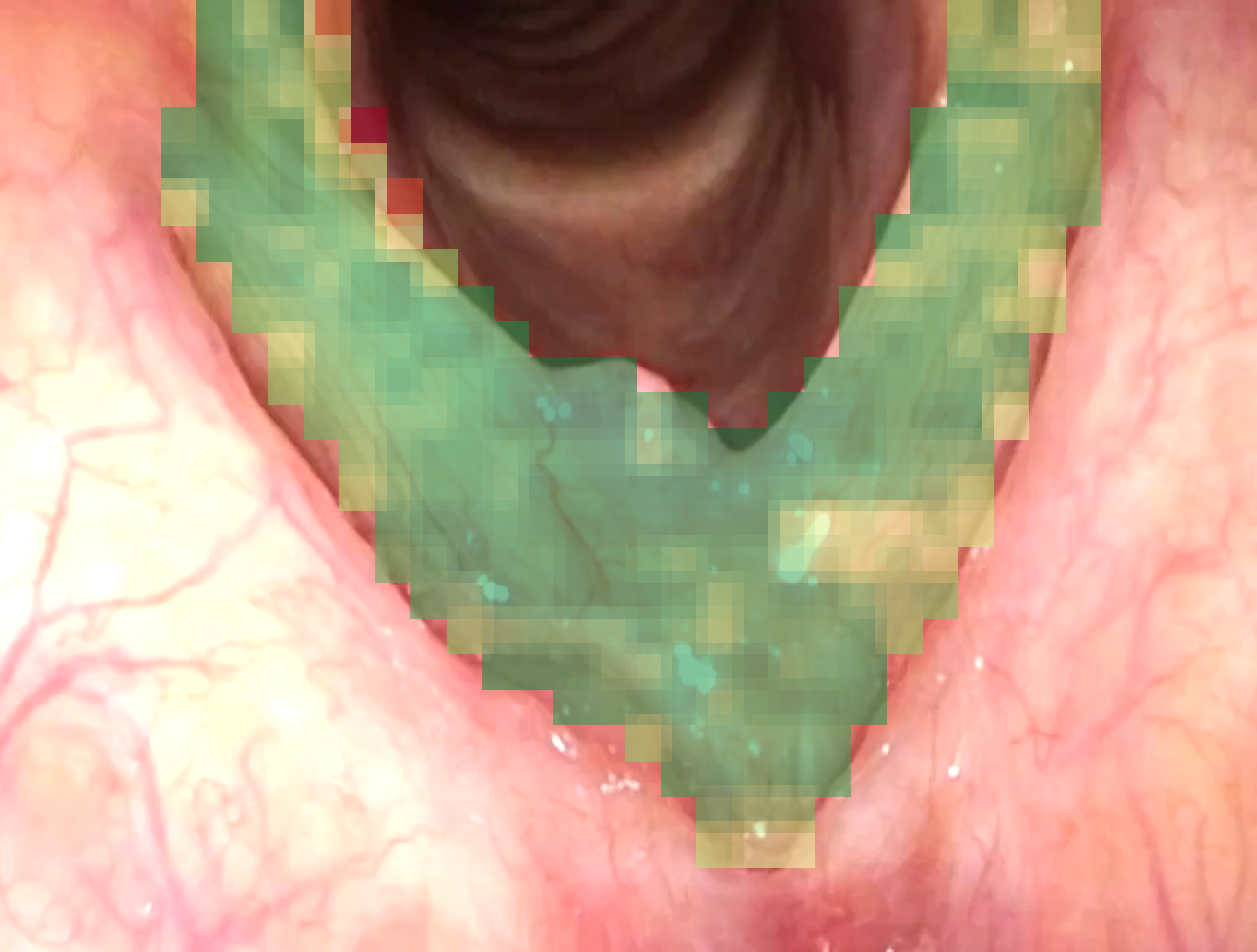}
        \caption{VF ROI in WLI; $\mathrm{mFBE_{tc}} \in [1,20]$ px.}
        \label{fig:vf_wli}
    \end{subfigure}

    \caption{Trackability heatmaps for VF and MPVP ROIs tracked with CoTracker3 using $\tilde{n}=300$ tracking candidates. Each heatmap was computed from a 4 s subvideo. Dark green and dark red represent the minimum and maximum $\mathrm{mFBE_{tc}}$ values, respectively, while the remaining colors represent intermediate values.}
    \label{fig:example_img}
\end{figure}

We evaluated two online point trackers, CoTracker3-Online~\cite{cotracker3} (reported as CoTracker3 in the tables) and TrackOnR~\cite{trackonr}, and the box tracker CSRT as a baseline~\cite{csrt}.

Table~\ref{tab:mfbe-scores} summarizes mFBE. Across all trackers and both ROI categories, NBI yielded lower mFBE than WLI, indicating higher forward--backward tracking consistency. For MPVP, the reduction was observed for all trackers but did not reach statistical significance. For VF, CoTracker3 and CSRT showed significant reductions in NBI ($p=0.013$ for both), while TrackOnR followed the same trend without reaching significance ($p=0.084$).

Table~\ref{tab:fr-scores} reports FR. CoTracker3 and TrackOnR showed low FR in both modalities, with no significant modality differences. CSRT had substantially higher FR overall. However, for VF, CSRT showed a significant FR reduction in NBI compared with WLI ($61.0\% \pm 24.0\%$ vs. $78.1\% \pm 15.7\%$; $p=0.001$), whereas FR was nearly identical between modalities for MPVP.

Qualitative heatmaps are shown in Figure~\ref{fig:example_img}. In the selected examples, NBI shows more contiguous low-error regions, whereas WLI appears more heterogeneous. These observations are consistent with the lower mFBE values observed for NBI, particularly for VF.

%% file: 04_discussion.tex
This study investigated whether imaging modality influences the trackability of natural visual landmarks in laryngeal endoscopy. The results showed that both NBI and WLI support tracking, while NBI generally provided more favorable conditions for forward--backward tracking consistency, with lower mFBE values across all trackers and ROI categories. The effect was strongest for the VF, where CoTracker3 and CSRT showed significant reductions, and TrackOnR showed the same trend without reaching statistical significance. FR remained low for CoTracker3 and TrackOnR in both modalities, while CSRT showed substantially higher FR overall but improved significantly in NBI for VF.

These findings are consistent with the hypothesis that NBI may provide more distinctive visual features through enhanced visualization of vascular and mucosal structures. This was also reflected in the trackability heatmaps (Figure~\ref{fig:example_img}), where NBI showed larger contiguous regions of low tracking error and more homogeneous consistency patterns. Importantly, the low FR of CoTracker3 and TrackOnR in both modalities suggests that tracking is feasible in both NBI and WLI when robust point trackers are used. This supports the perspective that landmark tracking could remain applicable in workflows where clinicians switch between modalities, although re-initialization may be required after the switch because image appearance changes substantially between WLI and NBI.

The benefit of NBI differed between ROI categories. VF tracking improved most clearly, whereas MPVP showed larger variability and no statistically significant modality differences. This may be explained by the more heterogeneous appearance of MPVP ROIs and their susceptibility to overexposure, specular reflections, motion blur, or local tissue changes. 

This study has several limitations. The analysis is based on forward--backward consistency rather than dense anatomical ground truth; therefore, low error indicates temporal self-consistency but not necessarily anatomical correctness. Future work should include manual landmark annotations, expert trajectory review, or sparse reference points to verify anatomical accuracy. In addition, the limited dataset size may reduce statistical power, especially for MPVP, where variability was high. Finally, NBI and WLI cannot be recorded simultaneously with the same endoscopic view. Although both modalities can be acquired from the same patients, future studies should follow predefined examination protocols to obtain more comparable NBI and WLI segments within each patient.

Overall, the findings suggest that NBI can improve the robustness of natural landmark tracking in laryngeal endoscopy, particularly for the vocal folds. At the same time, tracking remains feasible in both modalities when robust trackers are used, supporting the development of modality-aware tracking systems for NBI and WLI examinations.